\documentclass[11pt]{article}
\usepackage{graphicx}

\def\SDiff{S\kern-1.5pt di \kern-1pt f \kern-1.5pt f}
\def\ltsima{$\; \buildrel < \over \sim \;$}
\def\simlt{\lower.5ex\hbox{\ltsima}}
\def\gtsima{$\; \buildrel > \over \sim \;$}
\def\simgt{\lower.5ex\hbox{\gtsima}}

\begin{document}
\begin{center}
{\bf Elliptic CMB Sky}
\end{center}

\vspace{0.2in}

\noindent V.G.Gurzadyan$^{1,2}$, P. de Bernardis$^3$, G. De
Troia$^3$, C.L.Bianco$^2$, A.L.Kashin$^1$, H.Kuloghlian$^1$,
S.Masi$^3$, F.Piacentini$^3$, G.Polenta$^3$, G.Yegorian$^{1}$

\vspace{0.2in}

$^1$ Yerevan Physics Institute, Yerevan, Armenia; $^2$ ICRA, Dipartimento
di Fisica, University La Sapienza, Roma, Italy; $^3$ Dipartimento
di Fisica, University La Sapienza, Roma, Italy  \vspace{0.2in}

{\bf Abstract} - The ellipticity of the anisotropy spots of the
Cosmic Microwave Background measured by the Wilkinson Microwave
Anisotropy Probe (WMAP) has been studied. We find an average
ellipticity of about 2, confirming with a far larger statistics
similar results found first for the COBE-DMR CMB maps, and then
for the BOOMERanG CMB maps. There are no preferred directions for
the obliquity of the anisotropy spots. The average ellipticity is
independent of temperature threshold and is present on scales both
smaller and larger than the horizon at the last scattering. The
measured ellipticity characteristics are consistent with being the
effect of geodesics mixing occurring in an hyperbolic Universe,
and can mark the emergence of CMB ellipticity as a new observable
constant describing the Universe. There is no way of simulating
this effect. Therefore we cannot exclude that the observed
behavior of the measured ellipticity can result from a trivial
topology in the popular flat $\Lambda$-CDM model, or from a
non-trivial topology.

\section{Introduction}

Several properties of Cosmic Microwave Background (CMB) radiation
are known to contain crucial cosmological information. CMB maps
are among them. Recently, a significant ellipticity ($\simgt 2$)
of the anisotropies has been measured \cite{ellipse} in the high
signal to noise ratio CMB maps obtained at 150 GHz in the
BOOMERanG 1998 experiment \cite{Bern1}, \cite{Nett2002},
\cite{Bern2}, \cite{Ruhl}. The same effect was shown to exist at
scales larger than the horizon at the last scattering epoch
\cite{Gurz1}. The effect was confirmed when the WMAP \cite{Ben03}
data in the same region  were used \cite{Gurz2}, confirming the
good consistency of the BOOMERanG and WMAP CMB data \cite{Bern3}.
Ellipticity had been reported earlier for the COBE CMB maps
\cite{GT}.

However, the sky region observed by BOOMERanG was small (about 2\%
of the sky), and hence the statistics of anisotropy areas (spots)
was moderate; moreover the COBE map had neither high resolution
nor high signal-to-noise ratio. The WMAP first-year maps
\cite{Ben03}, instead, provide both. In this paper we analyze the
ellipticity in the WMAP W-channel maps of the full sky, and
confirm our previous estimates with higher significance, thus
firmly establishing the presence of an ellipticity effect.

The analysis of the ellipticity of the areas in CMB maps was
motivated by the effect of geodesic mixing \cite{GK1} (see
\cite{Pen} for a more general discussion) which has to occur in a
hyperbolic Universe due to the exponential divergence $L(t)$
of close null geodesics in (3+1)-space at the expansion of the scale factor $a(t)$
from its initial value $a(t_0)$
$$
   L(t)=L(t_0)\frac{a(t)}{a(t_0)}\exp(h\lambda),
$$
where $h$ is the Kolmogorov-Sinai entropy of the geodesic flow and
$\lambda$ is the affine parameter. What we denote as ellipticity
is actually the first order approximation of the distortion of the
spot. This is due to the exponential instability of the bundle of
geodesics, which leads to complex shape anisotropies, with
elongated amoeba-like tails. A more informative descriptor of such
structures is the Kolmogorov complexity \cite{G}.

Therefore, in our analysis, the numerical algorithms did not aim
to fit the given spot with an ellipse, but rather to define the
effective elongation via the ratio of two semi-axes of the spot.
This procedure, described in detail in \cite{ellipse}, implies the
estimation of Lyapunov exponents of the dynamical system defining
the effect, e.g. of the geodesic flow. The information on the
Lyapunov exponents and hence on Kolmogorov-Sinai entropy will
describe a dynamical system other than the geodesic flow, if being
responsible for the observed elongations.

Our analysis of WMAP maps strongly confirmed the existence of a
threshold independent mean ellipticity, in the range 2.1 - 2.5
(depending on the size of the areas i.e. on the number of the
pixels contained), over the whole interval of temperature
thresholds where the areas are well defined, and have reasonable
statistics.

We confirmed the existence of the effect for scales smaller and
larger than the horizon at recombination. This suggests that the
observed effect is not determined by the conditions at the last
scattering surface. Note also that for larger (several degrees)
areas, the biases of the estimator and of the noise are
negligible.

Numerical simulations of the CMB maps in the $\Lambda-CDM$ model
fitting the WMAP and Boomerang power spectra led to an average
ellipticity around 1.8 \cite{ellipse} for the noise level of
BOOMERanG. For the once popular CDM model, with no noise, the mean
ellipticity is around 1.4 \cite{Bond}.

Our determination of the ellipticity of the CMB sky makes stronger
the case that we could be dealing with the effect of geodesic
mixing in hyperbolic space. It is known that the low quadrupole
detected by WMAP \cite{Ben03} can also be explained by non-zero
curvature models, e.g. \cite{Efs03,Lum03,Aur03,Aur04}. However,
for a given curvature, an infinite number of topologies is
possible. This topological degeneracy excludes the possibility of
quantitative comparison of the observational data with a concrete
model, since any other model could fit even better. This situation
will persist until independent constraints on the topology will be
obtained, allowing to remove the degeneracy.

\section{Analysis}

We used the 94 GHz (3.2mm) maps from WMAP, because these
radiometers feature the highest angular resolution (beam size
0.22$^o$ FWHM) and also this band is least influenced by the
synchrotron radiation from our Galaxy.

We combined the data from the four independent detectors $w_i$
into two independent channels $A =(w_1+w_2)/2$ and $B =
(w_3+w_4)/2$.  We used Healpix \cite{Gorski} maps with nside=512
(6.9 arcmin pixel side), excluding all the pixels with galactic
latitude $|b|<20^{\circ}$. The ellipticity analysis was carried
out on the $A+B$ map, and we used the $A-B$ map to investigate the
effects of noise (see below).

The algorithms for the definition of the excursion sets have been
discussed in detail in previous papers \cite{ellipse, Gurz1}.
These algorithms define the hot areas with temperature equal and
higher than a given temperature threshold, and the cold areas with
temperature equal and lower than the threshold. The center, the
semi major and minor axes of the spots were determined by the
procedure described in \cite{ellipse}. We tested alternative
procedures obtaining consistent results. Since here we are dealing
with full-sky maps, we do not project the data on a plane, as was
justified in the analysis of small sky patches. We rather perform
our analysis on the sphere in Healpix representation.

In Figure 1 we present the average ellipticity (over the number of
areas) as a function of the temperature threshold for the $A+B$
map. The computation of the mean ellipticities over the considered
temperature intervals yields: $2.50\pm 0.03$, $2.18\pm 0.03$ and
$2.12\pm 0.05$ for areas containing more than 20, 50 and 100
pixels, respectively.

The statistics of the areas (detailed in Figure 2) is one to
two order of magnitudes higher than for the BOOMERanG data
analysis. However, the results shown in fig.1 are very consistent
(compare to \cite{ellipse}), thus confirming the absence of
instrumental / systematic effects.

The error bars shown in fig.1 are statistical only, as they are
computed from the standard deviation of the ellipticities of all
the areas at a given threshold.

We considered only the threshold interval where the areas are well
defined: at small thresholds the areas merge to one-connected
region, at higher thresholds the statistics becomes poor. In the
considered interval the behavior basically independent of
threshold for the ellipticity is evident.

No preferred direction can be defined for the obliquity of the
areas, as can be seen from the example in Figure 3. Here we have
selected areas containing from 100 up to 500 pixels in the two sky
belts at $b=\pm (20^o-40^o)$. The threshold is +200 $\mu K$. Such
constraints ensure having a sample for which the accuracy of the
obliquity estimator is homogenous. Similar results are obtained
for other area sizes and thresholds.

We also compared the results found selecting only the northern or
the southern hemisphere. No asymmetry of the ellipticity
descriptor is found (Figure 4).

We have shown in \cite{ellipse} that for areas containing
more than several tens of pixels, the algorithm bias does not
exceed 0.1.

Noise in the WMAP maps at 94 GHz is not negligible. The following
procedure was used to check its role in our ellipticity estimates.
New maps were obtained by adding the difference data $A-B$ (which
contain only noise) to different combinations of the $w_i$ data
(which contain signal and noise). This procedure effectively
changes the signal to noise ratio (S/N) of the map. We analyzed
the following maps: $A+B$; $(A'+B)/2$, where $A' =
(w_1+A-B+w_2)/2$; $w_1$; $w_1+A-B$. The normalized N/S ratio for
these maps is ($0.5$; $0.56$; $1.0$; $1.71$) respectively. We
apply to these new maps the same ellipticity analysis used for the
original $A+B$ map. The resulting mean ellipticities at a given
threshold are plotted in Figure 5. It is evident that there is in
fact a noise bias, making the measured ellipticity higher than the
true sky ellipticity. This bias is stronger for small spots, which
is natural. A naive linear extrapolation to zero noise would
suggest that the noise bias in the $A+B$ map is of the order of
0.1. This is an extrapolation, since we do not know the real law
describing the noise bias dependence on N/S. However, it is
absolutely natural to expect that the extrapolation curves flatten
for $N/S \ll 1$.  At S/N=4 the extrapolated ellipticities would be
2.3, 2.03, 1.93 for 20, 50 and 100-pixel areas. These can be
considered as lower limits. A maximum bias of 0.1 was obtained
also with numerical studies of the stability of the semi-axes and
of the ellipticity, when distorting only several pixels in areas
larger than several tens of pixels (see also
\cite{ellipse,Gurz1}).

We did not use simulated maps for the evaluation of the role of
the noise. The reason is that for producing simulated maps we
start from a model featuring an ellipticity parameter which can be
completely different from the real one. The ellipticity estimator
can depend on the noise in a way different than in the real case.
In this situation either independence or strong dependence of the
results on the noise can be misleading. This situation is well
known for many-parameter coupled systems. And is similar to the
situation we have for the topology, which is also usually
neglected in the list of parameters describing the cosmological
model. The role of the noise cannot be traced unambiguously unless
the maps are created for the given topology.

Therefore, we used the procedure described above to evaluate the
role of the noise based on the available real data. Obviously,
higher S/N data, e.g. the forthcoming 4-years WMAP maps or the B2K
maps, will allow to increase the accuracy of the above evaluation.

\section{Conclusions}

The spots in the CMB maps of WMAP have an ellipticity around 2. In
the previous analysis we could only state the compatibility of the
ellipticity with the threshold independence, either due to the low
statistics or to the biases. From the present analysis we have a
fair, robust establishment of such a behavior. This conclusion is
valid for small scales (say 20-pixels), as well as for larger
scales, with angular dimensions exceeding one degree.

The measured behavior of the ellipticity is compatible with the
following physical interpretation of the data. The temperature
threshold independence, the horizon-scale independence, and random
orientation are all predicted by geodesic mixing \cite{GK1}.
Moreover, from fig.5 there is an indication that the decrease in
the value of the ellipticity when increasing the size of the areas
is also intrinsic to the sky. In fact, the lines describing the
average ellipticity vs. N/S do not tend to converge at low N/S.
This behavior is also predicted by geodesic flows mixing
\cite{dynsys} in hyperbolic spaces which, locally, if the space is
not compact, behave as Anosov systems \cite{Anosov,LMP}. Again,
since there is no way of simulating this effect, we cannot exclude
that the observed behavior of ellipticity can result from a
trivial topology in the popular flat $\Lambda$-CDM model, or from
a non-trivial topology.

This analysis can mark the emergence of CMB ellipticity as a new
observable parameter describing the Universe, which has not been
considered in the usual list of parameters defining the
concordance cosmological model (see e.g. \cite{Ben03}).

As the power spectrum is degenerate with respect to ellipticity
and topology, future studies should aim to break such
degeneracies.

\bigskip
\bigskip

\begin{figure}[htp]
\begin{center}
\includegraphics[height=8cm,width=12cm]{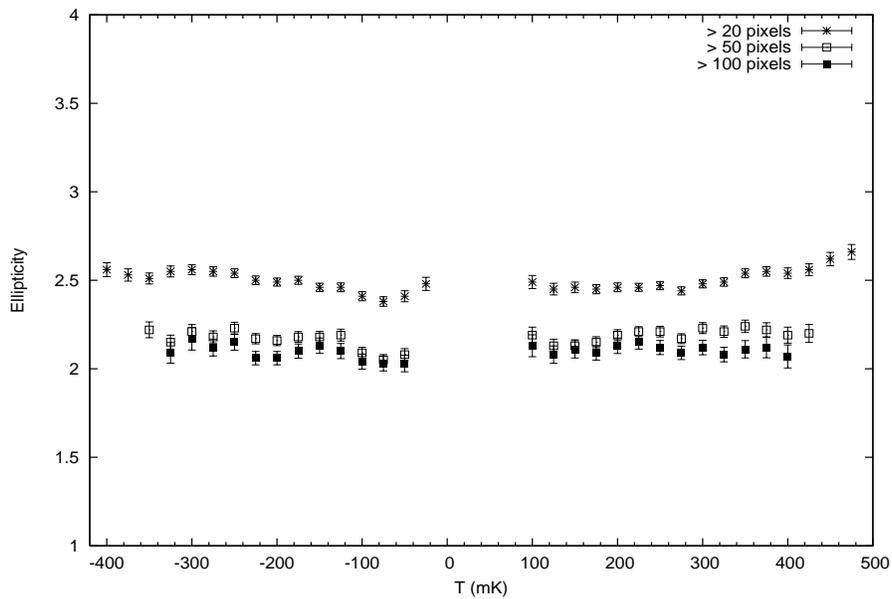}
\caption{Ellipticity vs temperature threshold (in $\mu K$)  for
the anisotropy areas containing more than 20, 50 and 100 pixels in
the 94 GHz sum map A = (w1+w2)/2 and B = (w3+w4)/2 of WMAP. The
error bars are statistical only. }
\end{center}
\end{figure}

\begin{figure}[htp]
\begin{center}
\includegraphics[height=8cm,width=12cm]{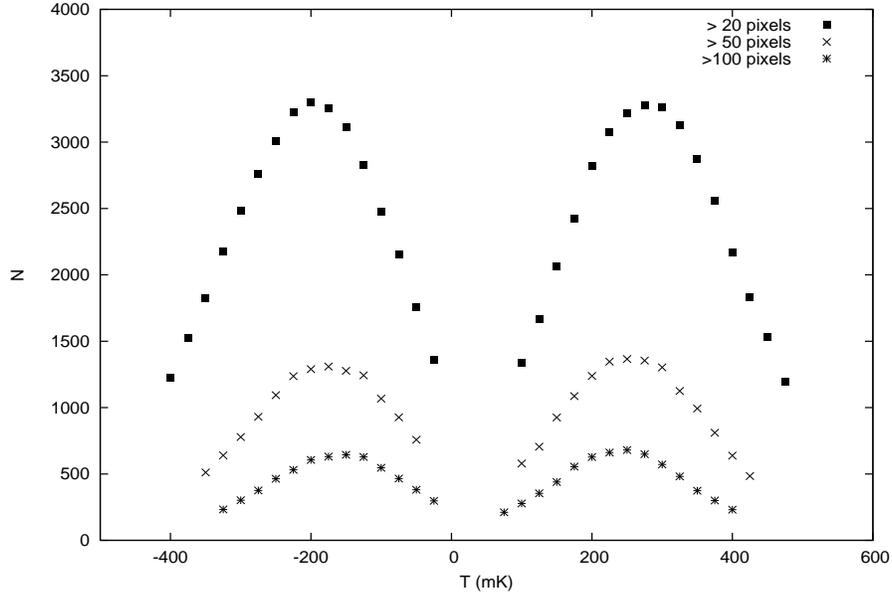}
\caption{The number of areas containing more than 20, 50 and 100
pixels vs the temperature threshold.}
\end{center}
\end{figure}

\begin{figure}[htp]
\begin{center}
\includegraphics[height=6cm,width=9cm]{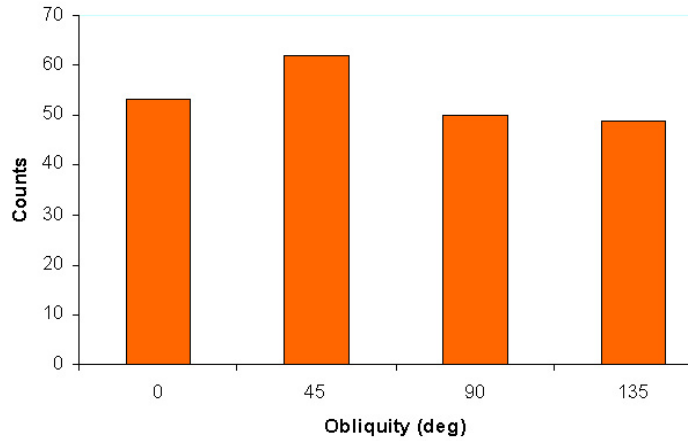}
\caption{Histogram of obliquities for anisotropy areas containing
100-500 pixels, at temperature threshold +200 $\mu K$ within the
two sky belts $\pm (20-40)^o$.}
\end{center}
\end{figure}

\begin{figure}[htp]
\begin{center}
\includegraphics[height=8cm,width=12cm]{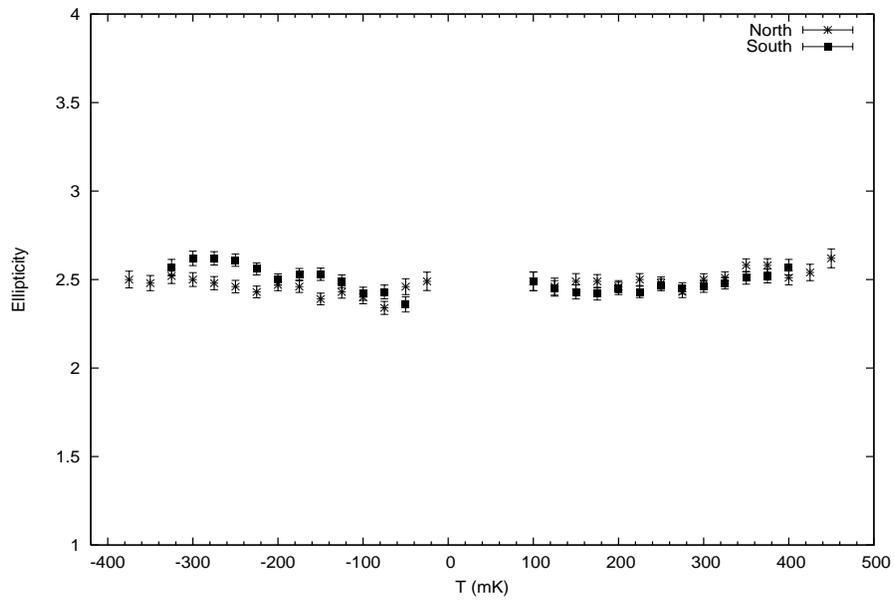}
\caption{Ellipticity of areas with more than 20-pixels of Southern
and Northern Galactic hemispheres.}
\end{center}
\end{figure}

\begin{figure}[htp]
\begin{center}
\includegraphics[height=8cm,width=12cm]{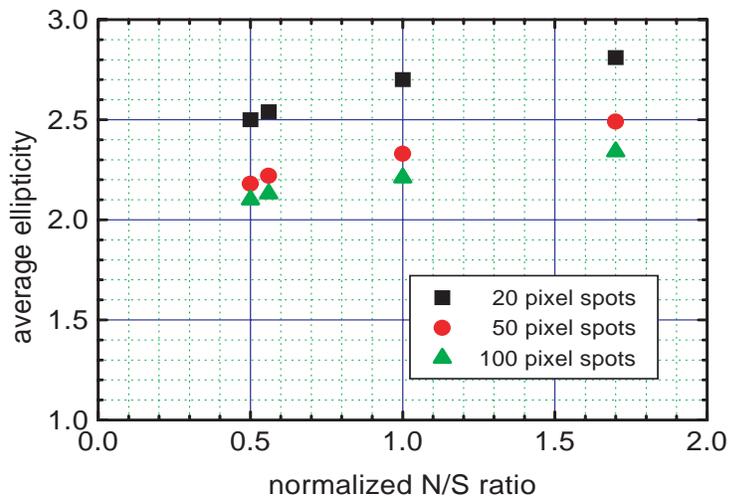}
\caption{Average ellipticity vs. Noise to Signal ratio. The noise
to signal ratio is normalized to be 1 for the map from detector
$w_1$ alone. This plot is used to investigate the noise bias on
the ellipticity estimate (see text).}
\end{center}
\end{figure}

\end{document}